          [1999/12/02 1.01 (PWD)] 
\documentclass[letterpaper,twocolumn]{esapub} 

\usepackage{natbib,graphicx} 
 
\title{Pair Production and Radiation Effects in Clouds Illuminated by Gamma Ray Sources} 
\author{C. D. Dermer} 
\affil{Code 7653, Naval Research Laboratory, Washington, DC 20375-5352, USA} 
\author{M. B\"ottcher\footnote{Chandra Fellow}} 
\author{E. P. Liang} 
\affil{Department of Space Physics and Astronomy, Rice University, Houston, TX 77005-1892, USA}

\newcommand{\psim}{\lower.5ex\hbox{$\; \buildrel \propto \over\sim \;$}} 
\def\gtrsim{\lower.5ex\hbox{$\; \buildrel > \over\sim \;$}}  
\def\lesssim{\lower.5ex\hbox{$\; \buildrel < \over\sim \;$}}  
  
\def\g2{\gamma_2}  
\def\tT{\tau_T} 
  
\def\e{{\epsilon}} 
 
\def\g{\gamma}

\begin{document} 
 
 
\maketitle 
 
\begin{abstract} 
Many classes of gamma-ray sources, such as gamma-ray bursts, blazars,
Seyfert galaxies, and galactic black hole sources are surrounded by
large amounts of gas and dust. X-rays and gamma-rays that traverse
this material will be attenuated by Compton scattering and
photoelectric absorption. One signature of an intervening scattering
cloud is radiation-hardening by electrons that have been scattered and
heated by the incident radiation, as illustrated by a Monte Carlo
calculation. Compton scattering provides backscattered photons that
will attenuate subsequent gamma rays through $\gamma\gamma$
pair-production processes. We calculate the pair efficiency for a
cloud illuminated by gamma-ray burst radiation. An analytic
calculation of the flux of X-rays and gamma rays Thomson scattered by
an intervening cloud is presented. Illuminated clouds near GRBs will
form relativistic plasmas containing large numbers of
electron-positron pairs that can be detected within $\sim$ 1-2 days of
the explosion before expanding and dissipating.  Localized regions of
pair annihilation radiation in the Galaxy could reveal gamma-ray
sources embedded in dense clouds, or sites of past GRB explosions.
\end{abstract} 
 
\section{Introduction} 
 
If a central source of hard X-rays and soft gamma rays is surrounded
by Thomson-thick scattering clouds, then the electrons in the cloud
will be heated to the Compton temperature of the incident radiation,
and hard X-rays will be backscattered. Radiation that passes through
these hot scattering clouds will have their soft X rays scattered to
higher energies \citep{db00}. Soft gamma-rays that pass through the
cloud will be subject to pair-production through $\gamma\gamma$
interactions \citep{mt00,mrr01}. These pairs can annihilate to form a
broadened 0.511 MeV line within the hot relativistic plasma, or will
diffuse into the interstellar medium to produce sites of localized
annihilation radiation. The Compton-scattered radiation is detected
later at lower flux levels \citep{mad00}. Such systems provide
potential targets for {\it INTEGRAL}.
 
A simple generalization of standard compactness arguments
\citep{fab86} for an illuminated cloud characterizes the condition
when backscattered pair-production interactions must be
considered. The process of $\g\g$ pair production attenuation begins
to be important when $n_\gamma
\sigma_{\gamma\gamma}r \gtrsim 1$, where $n_\gamma \simeq L_\gamma/(4\pi r^2 
m_ec^3)$ is the number density of $\gamma$-ray photons with energies
$\gtrsim 1$ MeV, $L_\gamma$ is the $\g$-ray luminosity, $r$ is the
radius of the cloud, and $\sigma_{\g\g}$ is the pair production cross
section. Because $\sigma_{\g\g}\approx \sigma_{\rm T}$ near threshold,
a system is compact when $L_\g/r \gtrsim 4\pi m_ec^3/\sigma_{\rm T} =
4.6\times 10^{29}$ ergs s$^{-1}$ cm$^{-1}$. We generalize this result
to determine when a cloud, illuminated by a $\g$-ray source a distance
$d$ from the cloud, becomes compact. The compactness condition depends
on the width $\Delta x$ of the illuminating photon front.  If $\Delta
x \gg r$, then the number density of the scattered radiation field is
$n_\g \approx L_\g
\tau_{\rm T}/(4\pi d^2 c\cdot m_e c^2)$, whereas if $\Delta x \ll r$, then 
$n_\g \approx L_\g \tau_{\rm T}(\Delta x/r)/(4\pi d^2 c\cdot m_e
c^2)$.  Consequently, an illuminated cloud with $\tau_{\rm T} \lesssim
1$ is compact if
\begin{equation} 
L_\g \gtrsim ({d\over r})^2\; {4\pi m_ec^3\over \sigma_{\rm T}}\;({r\over 
\tau_{\rm T}})\; \cases{1\; , & for $\Delta x\gg r$ \cr 
r/\Delta x\, & for $\Delta x \ll r$.\cr}
\label{eq1} 
\end{equation} 
For GRBs which produce photon fronts $\Delta x \sim 10$-100 lt-s,
Thomson-thick clouds with radii $r = 10^{15}r_{15}$ cm located $\sim
10^{16}$ cm away from the GRB source will be compact when $L_\g
\gtrsim 10^{50}$ ergs s$^{-1}$.  Smaller gamma-ray powers are required
for persistent sources. In this paper, we focus on GRB/cloud
interactions, in view of evidence \citep{ama00} that such sources have
highly metal enriched clouds with large column densities in their
vicinity.
 
\section{Blast-Wave Cloud Interaction} 
 
A wave of photons impinging on a cloud located $10^{16} d_{16}$ cm
from the GRB source will photoionize and Compton-scatter the ambient
electrons to energies characteristic of the incident $\gamma$ rays
\citep{mt00,db00}. For a plasma cloud with a width of $r \cong 3\times
10^4 r_{15}$ lt-s, radiation effects must be treated locally.  Besides
making pairs through $\gamma\gamma$ attenuation with backscattered
photons, the nonthermal electrons and pairs will Compton scatter
successive waves of photons, thereby modifying the incident spectrum.
 
GRBs produce a temporally-evolving spectrum of nonthermal synchrotron
photons with energy $\e = h\nu/m_e c^2$ that can be parameterized by
the equation
\begin{equation}  
\Phi(\e) = {L_p (1+\upsilon/\delta)\over 4\pi d^2m_ec^2 \e_0^2} 
\;\big[{1\over (\e/\e_0)^{2-\upsilon}+
(\upsilon/\delta)(\e/\e_0)^{2+\delta}}\bigr]\;  
\label{Phi} 
\end{equation} 
where $\e_0 $ is the photon energy of the peak of the $\nu F_\nu$
spectrum, $L_p$ is the peak spectral power, and $\upsilon$ and
$\delta$ are the $\nu F_\nu$ spectral indices at energies $\e\ll \e_0$
and $\e\gg \e_0$ respectively.  The quantities $L_p = L_p(t)$ and
$\e_0 = \e_0(t)$ depend upon time $t$, and $\e_0 \sim 1$ during the
prompt $\gamma$-ray luminous phase of a GRB. As nonthermal synchrotron
photons from a GRB impinge on the atoms in the cloud, electrons will
be Compton-scattered by the high-energy radiation. The time scale for
an electron to be Compton-scattered by a photon is $t_T(s) \approx 15
(1+z) d_{16}^2 \e_0/ L_{50}$, assuming that all Compton scattering
events occur in the Thomson limit.  The Klein-Nishina decline in the
Compton cross section will increase this estimate by a factor of $\sim
1$-3, depending on the incident spectrum.  Most of the electrons in
the cloud will therefore be scattered to high energies during a very
luminous ($L_{50}\gg 1$) GRB, or when the cloud is located at $d_{16}
\ll 1$.
 
Electrons are Compton-scattered on the Thomson time scale to form a
hard spectrum that turns over at kinetic energies of $\gtrsim
500\times$min($1,\e_0^2$) keV.  For a GRB with $\e_0 \sim 1$, most of
the kinetic energy is therefore carried by nonthermal electrons with
energies of $\sim 500$ keV.  Successive waves of photons that pass
through this plasma will continue to Compton-scatter the nonthermal
electrons.  Only the lowest energy photons, however, will be strongly
affected by the radiative transfer because both the Compton scattering
cross section and energy change per scattering is largest for the
lowest energy photons.
 
\begin{figure} 
\centering 
\includegraphics[width=0.9\linewidth]{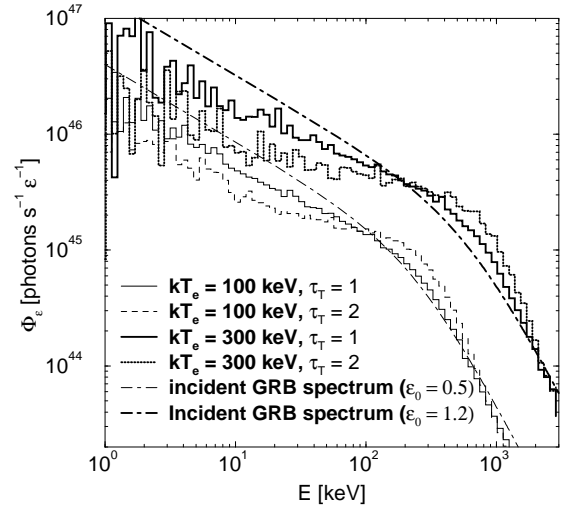} 
\caption{ 
Radiation transfer effects on GRB emission that passes through
electrons energized by an earlier portion of the photon front.  The
intrinsic spectra, from eq.(2) with $\upsilon = 2/3$ and $\delta =
0.2$, are shown by the thin and thick curves for $\e_0 = 0.5$ and
$\e_0 = 1.2$, respectively.  The nonthermal electrons are approximated
by a thermal distribution with temperatures of 100 keV (thin
histograms) and 300 keV (thick histograms), and Thomson depths $\tT =
1$ $\tT = 2$ as labeled.  Photon spectral indices $\alpha_X$
calculated at 50 keV are 0.54 ($T = 100$ keV, $\tT = 1$), 0.45 ($T =
300$ keV, $\tT = 1$), 0.28 ($T = 100$ keV, $\tT = 2$), and 0.16 ($T =
300$ keV, $\tT = 2$).
\label{fig1}} 
\end{figure}

Fig.\ 1 shows Monte Carlo simulations of radiation spectra described
by eq.\ (\ref{Phi}) that pass through a hot electron scattering
medium.  For simplicity, we approximate the hard nonthermal electron
spectrum by a thermal distribution with temperatures of 100 and
300~keV, and neglect pair production processes. The electron
temperature is chosen so that the mean electron kinetic energy is less
than the Compton temperature of the incident radiation.  These
calculations show that the lowest energy photons of the primary
synchrotron spectrum are most strongly scattered, and that the
``line-of-death" problem of the synchrotron shock model of GRBs
\citep{pea98,cl99} can be solved by radiation transfer effects through
a hot scattering cloud with $\tT \gtrsim$ 1-2.  Radiation transfer
effects by intervening clouds may likewise harden the intrinsic
spectrum of other hard X-ray and gamma-ray sources.

\section{Pair Efficiency Calculations} 
 
Following the initial wave of photons, successive photon fronts also
encounter the back-scattered radiation, which will lead to
$\gamma\gamma$ attenuation and pair production \citep{tm00}.  Here we
use the analytic model of \cite{dcb99}, eq.\ (\ref{Phi}), for the
spectral and temporal evolution of the GRB spectrum. The parameters
are total energy $E_0 = 10^{52} \, E_{52}$~ergs, initial bulk Lorentz
factor $\Gamma_0 = 300 \Gamma_{300}$, surrounding medium density $n_0
= 100 \, n_2$~cm$^{-3}$, radiative regime parameter $g = 1.6$, B-field
and electron-acceleration parameter $q = 2 \times 10^{-4}$, and we let
$\upsilon = 4/3$ and $\delta = 0.5$. The geometrical thickness of the
cloud is $r = 10^{15} \, r_{15}$~cm and its Thomson depth $\tau_T \sim
1$.  We calculate the space- and time-dependent $\gamma\gamma$ opacity
due to radiation backscattered by the cloud into the path of the
prompt radiation, and the corresponding rate of pair production.  For
this purpose, we assume that the cloud material is ionized instantly
at the onset of the GRB, and that the GRB radiation instantly heats
the free electrons to the Compton equilibrium temperature. The
formalism used to calculate the $\gamma\gamma$ opacity is largely
analogous to the calculation of \cite{bd95} for the case of
$\gamma\gamma$ attenuation in blazars. The effect of Compton
scattering on the transmitted GRB radiation is evaluated using an
appropriate average of the effective Compton scattering cross section
over the thermal distribution of electrons and pairs at any given
radius.
 
\begin{figure} 
\centering 
\includegraphics[width=0.9\linewidth]{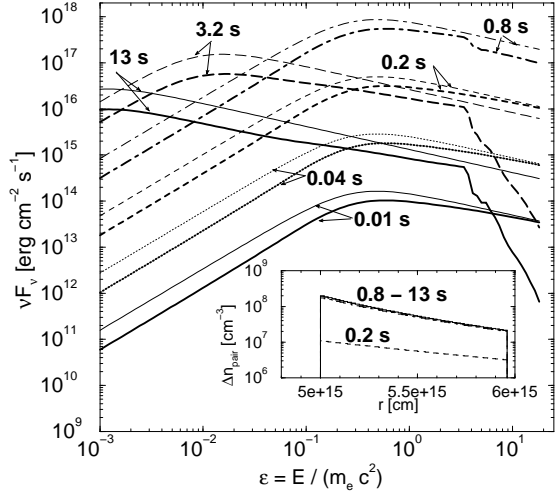} 
\caption{Spectral evolution of the pair and Compton attenuated GRB  
radiation. Inset shows the temporal evolution of the secondary pair  
density as a function of cloud radius. Parameters: $d_{16} = 0.5$, 
$r_{15} = 1$, $\tau_T = 1$, $E_{52} = \Gamma_{300} = n_2 = 1$. 
\label{fig2}} 
\end{figure} 
 
Fig.\ \ref{fig2} shows an example of such a calculation with typical
GRB parameters \citep{bd00}. The efficiency for making pairs in this
example (defined as the fraction of incident radiation energy
transformed into pairs) is 0.43 \%. This pair efficiency is large
enough that GRBs could make a substantial contribution to the
annihilation glow of the Milky Way, which would be concentrated near
sites of an earlier GRB. Fig.\ \ref{fig2} indicates that although the
actual pair efficiency and the relative number of pairs versus
background thermal electrons ($n_{\rm e, cloud} = 1.5 \times
10^9$~cm$^{-3}$) may be rather low, both the effects of $\gamma\gamma$
attenuation of high-energy photons and Compton upscattering of
low-energy photons can become quite significant.
 
\section{Scattered Flux from a Thomson-Thick Cloud} 
 
Here we outline a derivation of pulse scattering and echo effects that
occur when a photon front passes through a scattering cloud, with
Thomson depth $\tau_{\rm T} \lesssim 1$ \citep{mad00,bd00}, that lies
along the line-of-sight to the observer. Photons that are scattered by
an intervening cloud will be detected later at a reduced flux.
Following the treatment of \citet{bd95}, let $\dot
N_{ph}(\e,\Omega;t_*)d\e d\Omega$ represent the differential number of
photons with dimensionless photon energy $\e$ between $\e$ and $\e +
d\e$ that are directed into solid angle interval $d\Omega$ in the
direction $\Omega$ at time $t_*$ is the frame of the emitter. If, for
simplicity, we assume isotropic emission from the source, then $\dot
N_{ph}(\e;t_*)= 4\pi\dot N_{ph}(\e,\Omega;t_*)$. The photon emissivity
due to photons that are isotropically Thomson scattered by stationary
material with density $n_e({\bf r})$ at location ${\bf r}$ is $\dot
n_{ph}(\e ; {\bf r}, t_*) = n_e({\bf r})\sigma_{\rm T} \dot
N_{ph}(\e,t_*-r/c)/(4\pi r^2)$. The photon density received at a
location along the line of sight of the cloud at a distance $z$ from
the central source and at time $t$ is $n_{ph}(\e;z,t) = \int d^3 {\bf
r}\; \dot n_{ph}(\e; t-x/c)/(4\pi x^2 c)$, where $x = \sqrt{r^2 + z^2
- 2rz\mu}$, and $\mu$ is the cosine of the angle between the
directions to the received flux and the scattering event.
 
We assume that a scattering cloud lies along the line-of-sight between
the source and observer, and describe it with a special geometry
\begin{equation} 
n_e(r,\mu) = \cases{n_e^0\; , & for $r_1 
\leq r \leq r_2$ and $\mu_0 \leq \mu \leq 1$\cr 
0, & otherwise.\cr} 
\label{cloud} 
\end{equation} 
where $\sin\theta_0 = (r_2-r_1)/(r_2+r_1)$ and $\theta_0 =
\arccos(\mu_0)$. For a flare of constant intensity that lasts for a
time $t_1$, we obtain for the scattered flux
\begin{equation} 
\phi_{sc}(\e;z,t) = {n_e^0 \sigma_{\rm T} 
\dot N_{ph}(\e)\over 8\pi } \int_{\mu_0}^1 
{d\mu \over \mu} \; ({1\over z-R_2\mu} - {1\over z-R_1\mu})\;, 
\label{flux} 
\end{equation} 
where $R_1 =\max [r_1,(ct-ct_1-z)/(1-\mu)]$ and $R_2 = \min[r_2,
(ct-z)/(1-\mu)]$.
 
\begin{figure} 
\centering 
\includegraphics[width=1.0\linewidth]{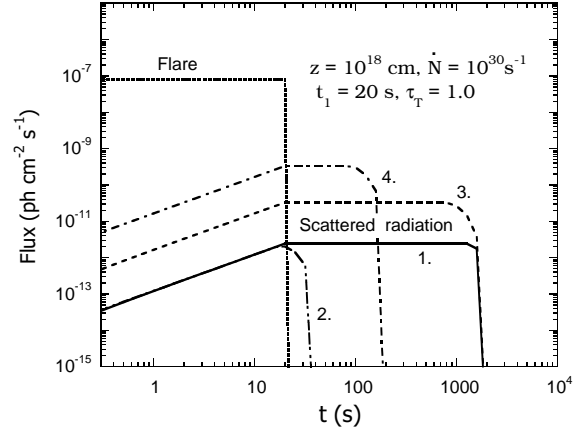} 
\caption{Direct and Thomson-scattered photon flux for a Thomson thick ($\tau_{\rm T} = 1$) cloud illuminated by a 20 second flare. The flux is normalized to a detector at $10^{18}$ cm from the source. Curve 1: $r = 10^{15}$ cm, $d =  10^{16}$ cm; Curve 2: $r = 10^{14}$ cm, $d= 10^{16}$ cm; Curve 3: $r_1 = 2.5\times 10^{14}$ cm, $d = 7.5\times 10^{14}$ cm; Curve 4: $r_1 = 2.5\times 10^{13}$ cm,$r_2 = 7.5\times 10^{13}$ cm. 
\label{fig3}} 
\end{figure} 
 
Fig.\ 3 shows calculations using eq.\ (\ref{flux}) for the direct and
scattered flux from a 20 s flare. The basic features of these
calculations can be understood by noting that the scattered flux
persists for a time
\begin{equation} 
t_{sc} = \max(t_1, {r^2\over 2 d c})\;, 
\label{tsc} 
\end{equation} 
and that the ratio of the scattered flux to the direct flux is  
\begin{equation} 
{\phi_{sc}\over \phi_{d}} \cong {c\over 2d}\;\tau_{\rm T} t_1\; 
,\label{phisc} 
\end{equation} 
for $\tau_{\rm T} \lesssim 1$. An interesting implication is that the
scattered flux is comparable in intensity to the direct flux only when
the scattering clouds are within a distance $\sim c t_1$ of the
central source. When the clouds are extremely optically thick with
$\tau_{\rm T} \gg 1$, the central source is strongly attenuated and
the scattered radiation will make a larger contribution.
 
\section{Observational Signatures of Illuminated Clouds} 
 
We have considered three potentially observable effects of gamma-ray
sources that illuminate nearby Thomson-thick clouds:
\begin{enumerate} 
\item Spectral hardening of radiation that passes through Compton-heated gas;  
\item Formation of pairs when $\gamma$ rays interact with backscattered photons; and 
\item Scattered flux that is detected at later times. 
\end{enumerate}  
We \citep{db00} have proposed that the first effect could account for
unusually hard spectra detected in $\sim 5$-10\% of GRBs, thereby
explaining exceptions to the nonthermal synchrotron shock model of
GRBs. {\it INTEGRAL} could observe such effects in serendipitous
GRBs. Given a full sky GRB rate of $\sim 900$ yr$^{-1}$, SPI might
detect $\sim 6$ GRBs yr$^{-1}$ within its fully coded FoV and $\sim
24$ GRBs yr$^{-1}$ within its partially-coded FoV. Searches for this
effect can be made through spectral analyses of these GRBs.
 
Production of e$^+$-e$^-$ pairs through the second effect will occur
in the vicinity of compact $\gamma$-ray sources that have large
amounts of gas surrounding them. The best prospects for INTEGRAL would
be to search for annihilation radiation from soft $\gamma$-ray and
unidentified EGRET sources in the disk of the Galaxy. GRBs will make a
significant contribution to the annihilation glow of the Galaxy. The
time-averaged kinetic energy of GRB sources into an L$^*$ galaxy such
as the Milky Way is $10^{40\pm 1}$ ergs s$^{-1}$ \cite{der00}. As
shown here, backscattered pair production processes can transform
$\sim 0.1$-1\% of the $\g$ rays into e$^+$-e$^-$ pairs with MeV
energies. Given the additional $\sim 10$\% efficiency for converting
the kinetic energy of GRBs into $\gamma$ rays, we see that GRB sources
produce a time-averaged injection rate of $\sim 10^{41}$-$10^{43}$
e$^+$ s$^{-1}$ into the Milky Way that could contribute substantially
to the Galactic 0.511 MeV intensity of the Galaxy (Dermer \& Murphy,
these proceedings). Because of the rarity and energy of GRBs, these
would most likely be detected as localized hot spots of annihilation
radiation surrounded by a hypernova remnant signaling an earlier GRB.
 
Compton echoes from distant scattering clouds may be detectable in the
afterglow phase of GRBs \citep{mad00}, and reverberation effects due
to Thomson scattering in heavily obscured Seyfert 2 galaxies could be
observable with {\it INTEGRAL}. As we have seen, scattered fluxes from
short pulses of high-energy radiation would be difficult to detect
from clouds in the vicinity of GRB sources unless $\tau_T \gg
1$. These clouds, would, however, be heated to temperatures of 100 keV
- several MeV to form relativistic plasmas that would expand and cool.
The cooling, expanding plasma would produce a broad pair annihilation
feature \citep{gs85}. Hot plasmas formed by nearby GRBs at $z \sim
0.1$ would be easily detectable with the {\it INTEGRAL} and Swift
missions \citep{db00}.
  
\section*{Acknowledgments} 
 
The work of C.D. is supported by the Office of Naval Research. The work of  
MB is supported by NASA through Chandra Postdoctoral Fellowship grant 
PF~9-10007, awarded by the Chandra X-ray Center, which is operated 
by the Smithsonian Astrophysical Observatory for NASA under 
contract NAS~8-39073.

\end{document}